\documentclass[fleqn,10pt]{wlscirep}
\usepackage[utf8]{inputenc}
\usepackage[T1]{fontenc}
\usepackage{lineno}
\usepackage{booktabs}
\usepackage{tabularx}
\usepackage{multirow}
\usepackage{float}
\title{A Chain-of-thought Reasoning Breast Ultrasound Dataset Covering All Histopathology Categories}

\author[1,$\dag$]{Haojun Yu}
\author[1,$\dag$]{Youcheng Li}
\author[2,$\dag$]{Zihan Niu}
\author[3,$\dag$]{Nan Zhang}
\author[4,$\dag$]{Xuantong Gong}
\author[6,$\dag$]{Huan Li}
\author[6,$\dag$]{Zhiying Zou}
\author[6,$\dag$]{Haifeng Qi}
\author[7,$\dag$]{Zhenxiao Cao}
\author[7]{Zijie Lan}
\author[7]{Xingjian Yuan}
\author[7]{Jiating He}
\author[7]{Haokai Zhang}
\author[7]{Shengtao Zhang}
\author[7]{Zicheng Wang}
\author[8]{Dong Wang}
\author[8]{Ziwei Zhao}

\author[6,*]{Congying Chen}
\author[4,5,*]{Yong Wang}
\author[3,*]{Wangyan Qin}
\author[2,*]{Qingli Zhu}
\author[1,*]{Liwei Wang}
\affil[1]{Peking University, School of Intelligence Science and Technology, Beijing, 100871, China}
\affil[2]{Peking Union Medical College Hospital, Beijing, 100730, China}
\affil[3]{Peking University Cancer Hospital \& Institute, Beijing, 100142, China}
\affil[4]{National Cancer Center / National Clinical Research Center for Cancer / Cancer Hospital, Chinese Academy of Medical Sciences and Peking Union Medical College, Beijing, 100021, China}
\affil[5]{The First Affiliated Hospital of China Medical University, Shenyang, 110001, China}
\affil[6]{Shenzhen Maternity \& Child Health Care Hospital, Shenzhen, 518028, China}
\affil[7]{Xi'an Jiaotong University, Xi'an, 710049, China}
\affil[8]{Yizhun Medical AI Co., Ltd, Beijing, 100000, China}

\affil[$\dag$]{These authors contributed equally to this work}

\affil[*]{corresponding author(s): Liwei Wang (wanglw@pku.edu.cn), Qingli Zhu (zhuqingli@pumch), Wangyan Qin (qinwangy@126.com), Yong Wang (wangyong@cicams.ac.cn), Congying Chen (chcoyi@qq.com)}

\begin{abstract}

Breast ultrasound (BUS) is an essential tool for diagnosing breast lesions, with millions of examinations per year. However, publicly available high-quality BUS benchmarks for AI development are limited in data scale and annotation richness.
In this work, we present BUS-CoT, a BUS dataset for chain-of-thought (CoT) reasoning analysis, which contains 11,439 images of 10,019 lesions from 4,838 patients and covers all 99 histopathology types.
To facilitate research on incentivizing CoT reasoning, we construct the reasoning processes based on observation, feature, diagnosis and pathology labels, annotated and verified by experienced experts.
Moreover, by covering lesions of all histopathology types, we aim to facilitate robust AI systems in rare cases, which can be error-prone in clinical practice.
The data and code are publicly available at \href{https://doi.org/10.6084/m9.figshare.29036876.v1}{https://doi.org/10.6084/m9.figshare.29036876.v1}.

\end{abstract}
\begin{document}

\flushbottom
\maketitle

\section*{Background \& Summary}

Breast cancer remains a significant threat to women's health, causing more than 670,000 deaths per year~\cite{xia2022cancer,siegel2023cancer,chhikara2023global,kim2025global}.
Accurate diagnosis of breast cancer based on medical images is crucial to improving prognosis. In developing countries, ultrasound has become an essential imaging tool for breast lesion diagnosis due to its cost efficiency, portability, noninvasiveness, and high sensitivity for younger women or dense breasts~\cite{gotzsche2013screening}. In China, more than 12 million breast ultrasound examinations are performed annually ~\cite{li2024breast}.

However, accurately interpreting breast ultrasound findings is challenging. For suspicious cases, the manual~\cite{guyatt1993users} recommends doctors to employ evidence-based \textbf{chain-of-thought (CoT) reasoning} — evaluating features like margins, echo patterns, and calcifications to estimate the probability of potential diagnoses. 
While breast ultrasound (BUS) AI systems have demonstrated remarkable success, they currently cannot provide this nuanced reasoning process. This limitation restricts their capacity to analyze challenging cases thoroughly.
Moreover, this lack of interpretability remains a significant gap in real-world applications. A single-blind randomized trial~\cite{goh2024large} revealed that although the AI achieved high diagnostic accuracy (92\%), diagnosticians assisted by AI saw only a marginal improvement in their diagnostic performance (from 74\% to 76\%). This discrepancy further highlights the importance of trustworthy chain-of-thought reasoning in human-AI interaction.

Another challenge is that AI systems face significant limitations in \textbf{out-of-domain (OOD) generalization} ~\cite{hong2024out}. Specifically, AI systems perform poorly when applied to categories absent in the training data.
Previous works~\cite{yu2024knowledge} ~\cite{tan20202019} demonstrate that histopathology is critical for BUS image analysis. Histopathology is determined by microscopic examination of biological tissues and reveals structural abnormalities and cellular changes which can be reflected by BUS semantics.
As the distribution of these categories is heavily long-tailed ~\cite{zhang2023deep}, those randomly collected datasets inevitably underrepresent or even exclude rare categories.
Consequently, AI systems trained on such datasets will exhibit substantial performance degradation in rare categories.
This limitation has critical implications for real-world practice, as rare categories account for a non-negligible proportion of patients — cases where diagnosticians are particularly prone to errors ~\cite{yang2022survey}.

\begin{figure}[h]
    \centering
    \includegraphics[width=0.7\textwidth]{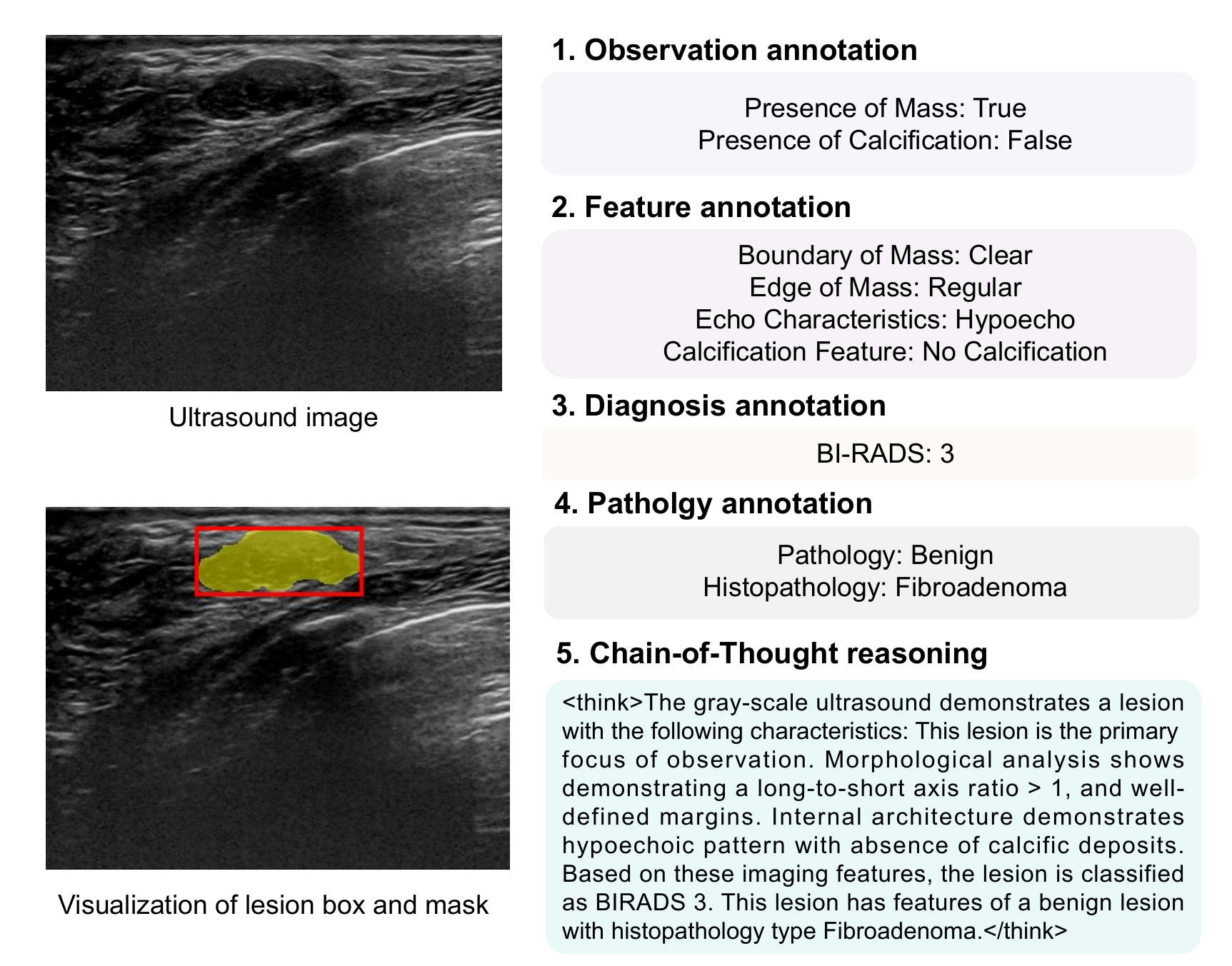}
\caption{A showcase of five-level annotations in BUS-CoT.}
    \label{fig_data_annotation}
\end{figure}

To address these challenges, we present the BUS-CoT dataset, a breast ultrasound dataset for CoT reasoning analysis, which contains 11,439 images of 10,019 lesions from 4,838 patients, covering all histopathology types~\cite{sinn2013brief}.
With recent advances in vision language models~\cite{hurst2024gpt,wang2024qwen2,guo2025deepseek}, AI systems now have the potential to perform CoT reasoning on medical images to analyse complicated cases~\cite{wei2022chain,miao2024chain}.
To facilitate research on incentivizing CoT reasoning~\cite{spak2017bi,mcdonald2016clinical}, we first annotated observation, feature, diagnosis and pathology labels, and then constructed the reasoning process based on these annotations, as shown in Fig\ref{fig_data_annotation}. The annotations are provided and verified by senior breast ultrasound experts with 8$\sim$26 years' experience.
Moreover, by including all histopathology categories in the BUS-CoT dataset, we provide an opportunity to develop robust AI systems without OOD generalization problems in terms of histopathology.
To further enhance the robustness, we provide an augmented version of BUS-CoT with 18 different device types using style transfer ~\cite{chu2017cyclegan}.

The contribution of this paper is three-fold. (1) We provide a large-scale high-quality BUS dataset, one order of magnitude larger than the mainstream benchmark BUSI~\cite{}. All images underwent rigorous quality control by experienced experts. (2) We provide CoT reasoning annotations and technical validation demonstrates that reasoning processes can enhance the model performance. (3) The BUS-CoT dataset covers all 99 histopathology categories and 18 device types, which has the potential to facilitate future works to address the OOD generalization problem.

\section*{Methods}

\subsection{Dataset characteristics}

BUS-CoT is an open multimodal dataset containing 10,897 records, including B-mode US, Doppler US, and Elastography. These records correspond to 10,019 lesions and 4,838 patients, categorized into different benign classes and malignant classes. Data labels contain lesion characteristics, US reports, BIRADS scores, and histopathology categories. These records make AI systems to replicate clinical reasoning from imaging features to pathological diagnosis. 
See Table~\ref{tab:stat_summary} for the statistics of our dataset. In this table we follow the principle of WHO~\cite{allison2019classification} in categorizing subtypes. 

\begin{table}[H]
\centering
\renewcommand{\arraystretch}{1.35}
\caption{{\bf Dataset Summary by Pathology.}}
\label{tab:stat_summary}
\begin{tabular}{ l | m{4.5cm} | m{4.5cm} | m{3cm} }
    \toprule
    \textbf{Pathology} & \textbf{Benign} & \textbf{Malignant} & \textbf{Others} \\
    \midrule
    Lesions & 4,856 & 4,814 & 349 \\
    Categories & {Fibroadenoma}  {(1,047)}   & {Invasive ductal carcinoma} {(896)} & {Others} {(349)} \\
    & {Phyllodes tumour}  {(74)}  &{Invasive lobular carcinoma} {(75)}  & \\
    & {Intraductal papilloma} {(72)}&{Ductal carcinoma in situ} {(72)}  & \\
    & {Atypical ductal hyperplasia} {(62)}& {Mucinous carcinoma} {(64)}&\\
    & {Radial scar} {(56)} &{Paget disease of the breast} {(57)}& \\
    & {Other benign} {(3,545)}& {Other malignant} {(3,650)}&\\
    \bottomrule
\end{tabular}
\end{table}

\subsection{Data collection}

\begin{figure}[h]
    \centering
    \includegraphics[width=\textwidth]{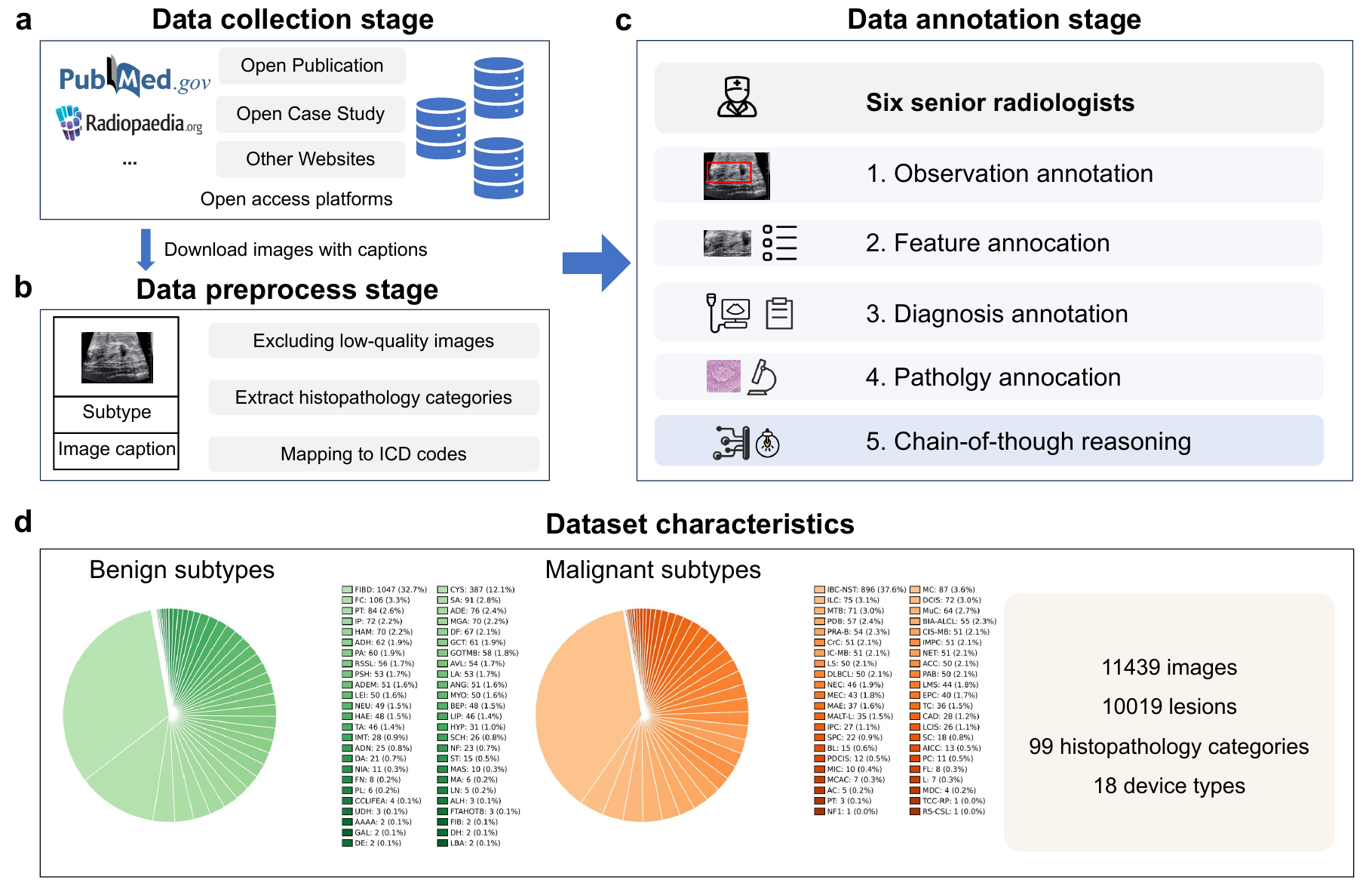}
\caption{
\textbf{BUS-CoT dataset.} 
\textbf{(a) Collection}: data collection from public available resources.
\textbf{(b)  Preprocessing}: Data cleaning and metadata extraction.
\textbf{(c)  Annotation}: observation, feature, diagnosis, pathology, chain-of-though reasoning annotations.
\textbf{(d)  Statistics}: distribution of histopathology categories.
}
    \label{fig_main}
\end{figure}

We collect data from open-access papers, publicly available case studies (like Radiopaedia and PubMed), and open-access datasets that contain biopsy results or clear pathological subtype annotations using histopathology categories as search terms as shown in Figure~\ref{fig_main} a.
We have designed strict data inclusion criteria to prevent contamination by dirty data as below. We collect data according to the breast cancer pathological categories as specified by the WHO~\cite{}. Our annotation team includes six senior radiologists and is reviewed by another independent physician team.

\subsection{Data preprocessing}

To ensure the sanity and integrity of the data, we performed the following preprocessing operations, see also Figure~\ref{fig_main} b. By excluding low-quality images, we aimed to ensure effectiveness and minimize redundancy. Ultrasound regions were cropped to avoid interruption from the background. From description texts, we also extracted histopathology categories to support medical prior knowledge and mapping them into ICD codes~\cite{} by expert manual annotation.

\subsection{Data annotation}

As illustrated in Figure~\ref{fig_main} c, our data annotation follows a rigorous three-stage annotation protocol. Six senior ultrasound physicians performed the initial annotation, adhering to standardized clinical guidelines. First, they conducted observation annotation by identifying lesion presence, localizing lesions, and documenting calcifications and echogenicity patterns. Next, they provided feature annotation through detailed morphological characterization, including shape categorization (e.g., oval, irregular), margin analysis (circumscribed vs. angular), density grading, and internal echo pattern classification. Finally, they completed diagnosis annotation by assigning BI-RADS scores and correlating imaging features with confirmed histopathology categories.

\section*{Data Records}

\begin{figure}[h]
    \centering
    \includegraphics[width=\textwidth]{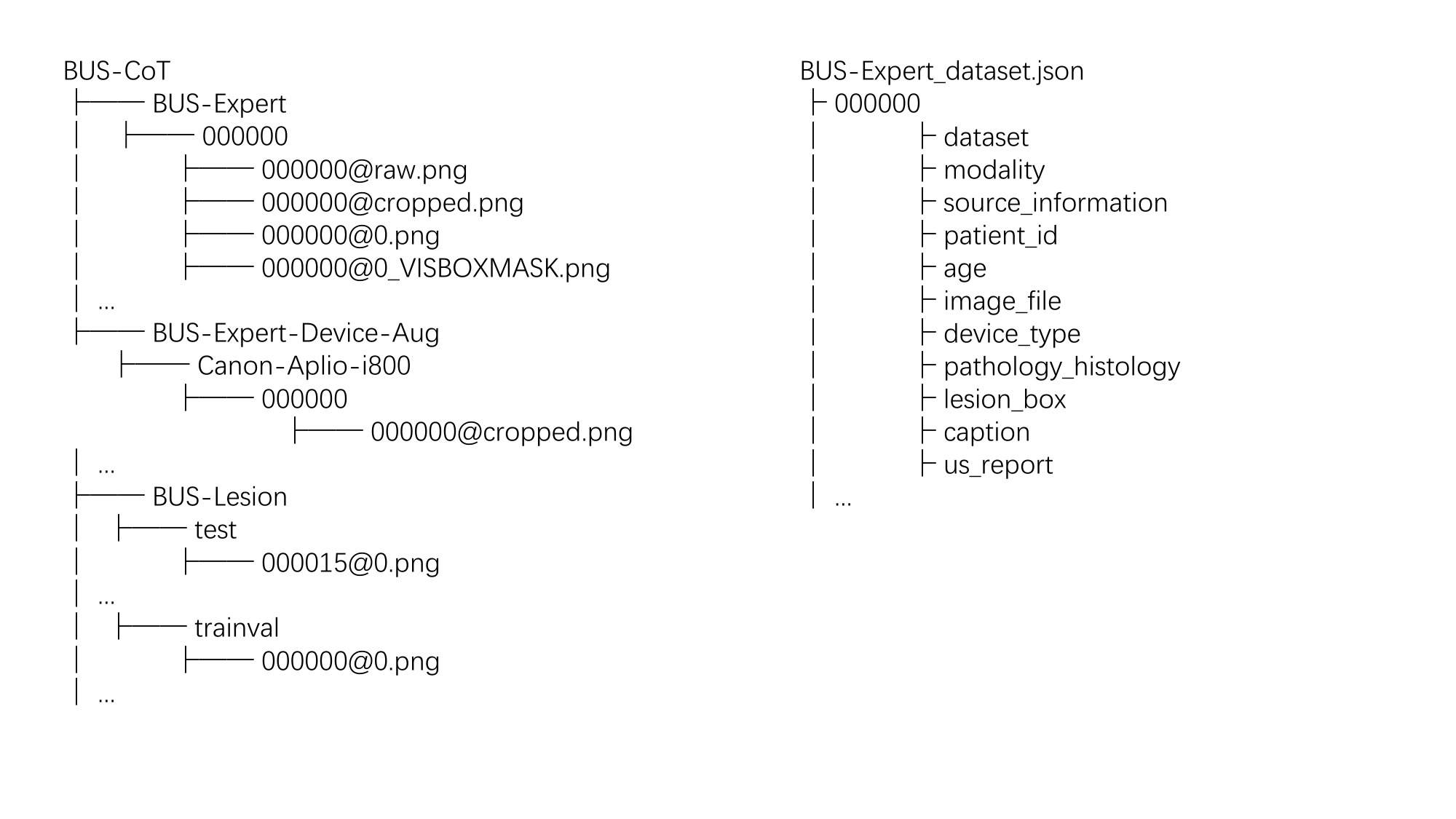}
\caption{
\textbf{Hierarchy of BUS-CoT dataset.} 
}
    \label{fig_data_str}
\end{figure}
The dataset is organized under a root directory containing an image folder and a separate JSON annotation file, with each record assigned a unique 6-digit identifier. The image folder includes four PNG-format files per case as illustrated in Figure \ref{fig_data_str}.

\section*{Technical Validation}

\begin{figure}[h]
    \centering
    \includegraphics[width=0.8\textwidth]{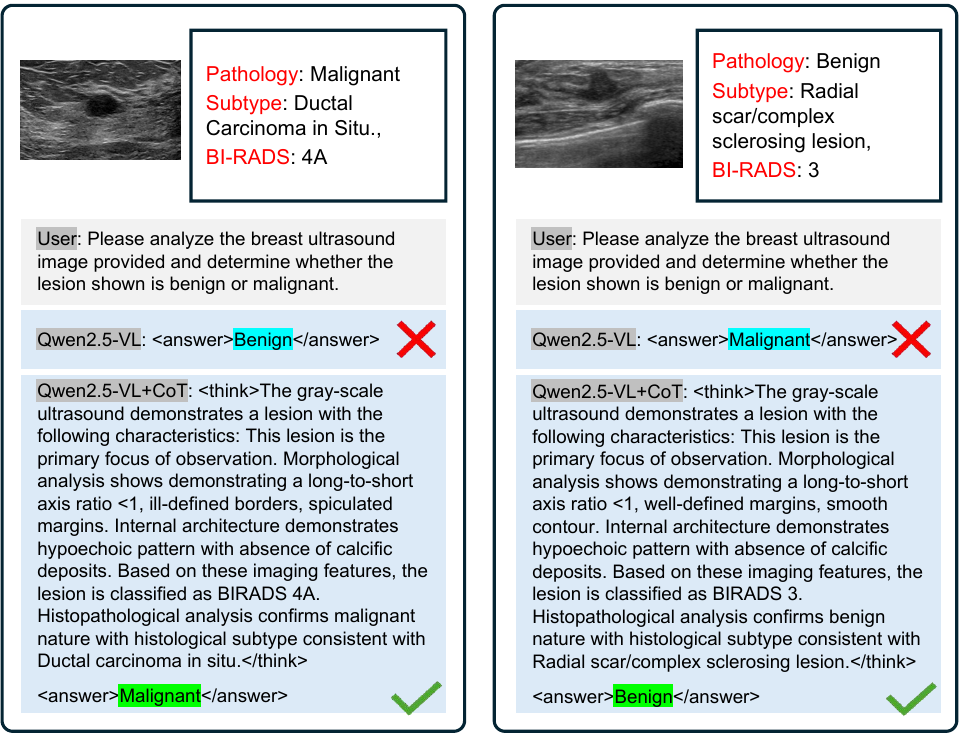}
\caption{Chain-of-Thought reasoning validation cases.}
    \label{fig_case_study}
\end{figure}

\begin{table}[!ht]
\centering
\caption{Performance Comparison on different approaches on our BUS-CoT dataset}
\label{tab:performance}
\begin{tabular}{l|ccc}
\toprule
\textbf{Model} & \textbf{AUROC} & \textbf{AUPRC} & \textbf{ACC} \\ 
\hline

ResNet 50  & 0.8085 \scriptsize{(0.7813, 0.8361)} & 0.7615 \scriptsize{(0.7190, 0.8068)} & 0.7382 \scriptsize{(0.7077, 0.7676)} \\
ResNet 101  & 0.8192 \scriptsize{(0.7944, 0.8464)} & 0.7737 \scriptsize{(0.7335, 0.8177)} & 0.7434 \scriptsize{(0.7150, 0.7708)} \\
ResNet 152  & 0.8183 \scriptsize{(0.7925, 0.8455)} & 0.7770 \scriptsize{(0.7376, 0.8232)} & 0.7466 \scriptsize{(0.7192, 0.7771)} \\
Swin-B  & 0.8371 \scriptsize{(0.8129, 0.8632)} & 0.7979 \scriptsize{(0.7613, 0.8359)} & 0.7476 \scriptsize{(0.7203, 0.7750)} \\
Swin-L  & 0.8458 \scriptsize{(0.8227, 0.8712)} & 0.7988 \scriptsize{0.7627, 0.8394)} & 0.7497 \scriptsize{(0.7224, 0.7771)} \\
ViT-B  & 0.8287 \scriptsize{(0.8043, 0.8551)} & 0.7930 \scriptsize{(0.7541, 0.8346)} & 0.7487 \scriptsize{(0.7213, 0.7771)} \\
ViT-L  & 0.8340 \scriptsize{(0.8094, 0.8603)} & 0.7819 \scriptsize{(0.7402, 0.8284)} & 0.7445 \scriptsize{(0.7161, 0.7718)} \\
Qwen2.5-VL-3B & - & - & 0.7447 \scriptsize{(0.7173, 0.7732)} \\
Qwen2.5-VL-3B + CoT & - & - & \textbf{0.7779 \scriptsize{(0.7514, 0.8044)}} \\
\bottomrule
\end{tabular}
\end{table}

The BUS-CoT dataset contains 10,897 ultrasound images spanning 10,019 lesions across 99 pathological categories, ensuring broad coverage of clinical scenarios—critical statistical characteristics, including subtype prevalence and malignancy distribution. To ensure clinical relevance and prevent data leakage, we implemented a patient-level 8:2 train-test split, rigorously isolating images from individual patients to either training or evaluation sets. This strategy mirrors real-world deployment conditions where models must generalize to unseen patient populations.

For conventional image classification, ResNet, Swin-Transformer, and ViT architectures were configured with global average pooling and fully connected layers for binary malignancy prediction. The Qwen2.5-VL framework was adapted to generate diagnostic dialogues through aspect-ratio-preserved $224\!\times\!224$ image inputs and cross-modal attention mechanisms. Low-Rank Adaptation (LoRA) with rank=64 and $\alpha=64$ enabled efficient fine-tuning while preserving pretrained knowledge, balancing parameter efficiency with diagnostic performance.

Classification models underwent five-fold cross-validation on zero-padded $224\!\times\!224$ images using AdamW optimization ($\beta_1=0.9$, $\beta_2=0.999$), batch size 16, initial learning rate $5\!\times\!10^{-5}$ (cosine decay), and weight decay 0.1 over 5 epochs. Qwen2.5-VL was fine-tuned across 16$\times$NVIDIA 4090 GPUs with gradient accumulation (step=4) over 10,000 steps at constant learning rate $5\!\times\!10^{-5}$. Chain-of-thought prompting significantly outperformed direct classification in ambiguous cases, improving AUC-ROC by 3\% for lesions with overlapping benign/malignant features.

\textbf{Limitations.} The current dataset lacks patients' clinical information, such as medical records, family history, chief complaints, blood tests, habits, palpation results, etc. These kinds of text information can be critical for clinical diagnosis~\cite{acosta2022multimodal}. The future work is to collect and integrate multimodal clinical data to enhance diagnostic accuracy and model interpretability.

\section*{Code availability}
The code used in this study was written in Python3.9 and is available. The code is based on PyTorch (version 2.4.0+cu121). The data and code are publicly available at \href{https://doi.org/10.6084/m9.figshare.29036876.v1}{https://doi.org/10.6084/m9.figshare.29036876.v1}.

\bibliography{sample}

\section*{Acknowledgements} 

This work is supported by National Science and Technology Major Project (2022ZD0114902) and National Science Foundation of China (NSFC62276005). We would like to acknowledge all authors of the open datasets used in this study.

\section*{Author contributions statement}

H.Y. and L.W. conceived the experiments. H.Y., Y.L., Z.N., Q.Z, D.W., Z.Zh. developed the code and benchmark. 
Z.C., Z.L., X.Y., J.H., H.Z., S.Z. and Z.W. contributed to data collection and cleaning.
Z.N., N.Z., X.G., H.L., Z.Zo., H.Q., C.C., Y.W., W.Q. and Q.Z. contributed to data annotations.
H.Y., Y.L. and Z.C. wrote the manuscript.
All authors reviewed the manuscript.
\section*{Competing interests} 

The authors declare no competing interests.
\end{document}